\documentclass[12pt]{article}
\usepackage[top = 2 cm, bottom = 3 cm, left = 2.5 cm, right = 1.5 cm]{geometry}
\usepackage{authblk, cite, color, amssymb, amsmath, graphicx}
\usepackage[colorlinks = true, linkcolor = blue, citecolor = red]{hyperref}
\usepackage[dvipsnames]{xcolor}

\begin{document}

\title{Black Hole Information Problem and Wave Bursts}

\author{{\bf Merab Gogberashvili$^{1,2}$} and {\bf Lasha Pantskhava}$^1$}
\affil{\small $^1$ Javakhishvili Tbilisi State University, 3 Chavchavadze Avenue, Tbilisi 0179, Georgia \authorcr
$^2$ Andronikashvili Institute of Physics, 6 Tamarashvili Street, Tbilisi 0177, Georgia}

\maketitle

\begin{abstract}

By reexamination of the boundary conditions of wave equation on a black hole horizon it is found not harmonic, but real-valued exponentially time-dependent solutions. This means that quantum particles probably do not cross the Schwarzschild horizon, but are absorbed and some are reflected by it, what potentially can solve the famous black hole information paradox. To study this strong gravitational lensing we are introducing an effective negative cosmological constant between the Schwarzschild and photon spheres. It is shown that the reflected particles can obtain their additional energy in this effective AdS space and could explain properties of some unusually strong signals, like LIGO events, gamma ray and fast radio bursts.

\vskip 1mm

\end{abstract}
\vskip 5mm


Gravitational wave (GW) signals \cite{GW}, gamma ray bursts (GRBs) \cite{GRB} and fast radio bursts (FRBs) \cite{FRB} are examples of extragalactic transients, whose origin and emission mechanisms are still to some degree unknown. Usually these events are associated with massive black holes (BHs), the most enigmatic objects in the universe. This article attempts to describe these unusually strong signals by means of the amplification of the gravitationally strongly lensed waves by BHs. This idea will explain repeat GWs, GRBs and FRBs detected from the same location and agrees with the opinion that these signals could be physically connected.

Weak gravitational lensing by a massive object is well studied \cite{Lens}, however, wave deflection still remains a difficult subject in the strong gravitational field \cite{Darwin}. Several attempts were made to obtain adequate lens equations for so called relativistic lensing in Schwarzschild space-time \cite{StrLens}.

The most promising focusing agents for any kind of waves are BHs, which are often located at the centers of galaxies and are surrounded by dust clouds. So there are problems with the observation of the BH lensings (due to spurious radiation and large extinctions of waves by accreting materials), especially of relativistic images, which should be formed closer to the BHs \cite{Virb}.

In the case of electromagnetic radiation it is expected that, due to absorptions (larger for smaller wavelength), only the short duration relativistic lensing signals in the forms of radio (largest wavelength) and gamma (most energetic) waves could escape the dust clouds surrounding the BH. For GWs, dust and noise do not present an obstacle, also the collinear requirement is less severe (since GWs occur at low frequencies) and the resulting focused region has a relatively large area (because of diffraction). These increase the chances to observe several relativistic images of the same GW source, since waves passing close to the BH can loop around it before reaching an observer. Then, on the same side as the source from the optical axis, a primary image (formed by weak lensing) and several relativistic images will be visible \cite{Virb}. So, unlike the case with the single GRBs and FRBs impulses, a distant observer will detect periodic GW signal.

In this article we reanalyze boundary conditions for wave equations at the BH horizon and show the existence of real-valued time-dependent and non-harmonic exponential solutions \cite{Gog}. This means that quantum particles probably do not cross the Schwarzschild horizon, but are absorbed or reflected by it. This potentially solves black hole information problem (see recent review \cite{Mar}). Moreover, reflected in certain conditions (small or zero masses and angular momentum of the infalling on a BH particles) waves from the Schwarzschild sphere could obtain energy from the gravitational field and create the burst-type signals that are observed.

Let's for the simplicity consider motion of quantum particles in the field of a non-rotating, uncharged BH with the Schwarzschild line element,
\begin{equation} \label{Schwarzschild}
ds^2 = f(r) dt^2 - \frac {dr^2}{f(r)} - r^2 (d\theta^2 + \sin^2 \theta d\phi^2)~,
\end{equation}
where
\begin{equation} \label{f}
f(r) = 1 - \frac {2M}{r}~. ~~~~~~ \left \{
\begin{array} {lr}
2M \leq r \leq \infty\\
0 \leq f \leq 1
\end{array}
\right.
\end{equation}
In our system of units, $c = \hbar = G =1$, the parameter $2M$ here corresponds to the Schwarzschild horizon and has the dimension of length.

Without loss of generality, close to the Schwarzschild horizon one can study massless scalar particles, since it is known that close to the horizon mass terms in the particle equations can be neglected and also in the eikonal approximation the polarization tensors are parallel-transported along the null geodesic and in many cases can be regarded to be constant \cite{Chandra}.

The wave (massless Klein-Gordon) equation in a curved space-time can be written in the form:
\begin{equation} \label{wave}
\Box \Phi = \frac {1}{ \sqrt{-g}}\partial_\mu \left( \sqrt{-g}g^{\mu\nu}\partial_\nu \right) \Phi = 0~.
\end{equation}
The metric (\ref{Schwarzschild}) is highly symmetric, so in the equation (\ref{wave}) we can separate the variables,
\begin{equation} \label{Phi}
\Phi \sim \psi (t, r) Y_{lm}(\theta, \phi) ~,
\end{equation}
where $Y_{lm}(\theta, \phi)$ are spherical harmonics. Then the equation (\ref{wave}) gives:
\begin{equation} \label{psi}
\left[ r^2 \partial_t^2 - f \partial_r\left( r^2 f\partial_r\right) + l(l+1) f\right]\psi (t,r) = 0~,
\end{equation}
where $l$ is the orbital angular momentum quantum number.

Let us assume that close to the horizon, $r \approx 2M$, it's possible to separate also the remaining variables in (\ref{psi}),
\begin{equation} \label{psi=RT}
\psi (t,r) = \frac 1r T(t)R(r) ~.
\end{equation}
Then (\ref{psi}) leads to the system of equations:
\begin{equation}
\partial^2_t T = C T ~, \label{T''}
\end{equation}
\begin{equation}
f^2 \partial_r^2 R + \frac{2Mf}{r^2} \partial_r R - \left\{C + f\left[ \frac {2M}{r^3} + \frac {l(l+1)}{r^2}\right]\right\} R = 0~, \label{R''}
\end{equation}
where $C$ is a separation constant.

To determine physical boundary conditions the equation (\ref{R''}) usually is transformed using the Regge-Wheeler 'tortoise' coordinate,
\begin{equation} \label{tortoise}
r^* = \int \frac {dr}{f} = r + 2M \ln \left( \frac {r}{2M} - 1\right)~, ~~~~~~ \left \{
\begin{array} {lr}
2M \leq r \leq \infty\\
-\infty \leq r^* \leq \infty
\end{array}
\right.
\end{equation}
which brings (\ref{R''}) to a Schr\"{o}dinger-like  equation:
\begin{equation} \label{KG-R}
\frac {d^2R}{dr^{*2}} - \left[V(r^*) + C\right] R = 0 ~,
\end{equation}
where the effective potential for the considered massless case is
\begin{equation} \label{Veff}
V(r^*) = f \left[\frac {2M}{r^3} + \frac {l(l+1)}{r^2}\right] ~.
\end{equation}

To set boundary conditions for the equation (\ref{KG-R}) close to the BH horizon ($r^* \to - \infty$, or $f \to 0$ and $V(r^*)\to 0$) commonly it is assumed the existence of a horizon crossing, ingoing radial waves \cite{Star,Matz},
\begin{equation} \label{solution-f=0}
T(t)_{f \to 0} \sim  e^{\pm i\omega t}~, ~~~~~ R(r)_{f \to 0} \sim e^{\pm i\omega r^*} ~,
\end{equation}
corresponding to the negative separation constant in the system (\ref{T''})-(\ref{R''}),
\begin{equation}
C = -\omega^2 < 0 ~.
\end{equation}
However, the transformation (\ref{tortoise}) is singular and the solution (\ref{solution-f=0}), due to appearance of the Dirac delta function in second derivatives of the 'tortoise' coordinate function (\ref{tortoise}), does not obey the equation (\ref{R''}).

To explain this point let us consider the model equation,
\begin{equation} \label{cal-R}
r(r-1)^2 \partial_r^2 {\cal R} + (r-1) \partial_r {\cal R} + r^3 {\cal R} = 0~,
\end{equation}
which has similar to (\ref{R''}) kinetic part, with $2M = 1$. It is obvious that in (\ref{cal-R}), due to the last term, for a regular solution one should use the boundary condition
\begin{equation} \label{}
{\cal R}|_{r\to 1} \to 0~.
\end{equation}
From the other hand, using the singular 'tortoise' coordinate,
\begin{equation} \label{y}
y = r + \ln \left( r - 1\right)~,
\end{equation}
the equation (\ref{cal-R}) can be transformed to the form:
\begin{equation} \label{cal-r}
\partial_y^2 {\cal R} + {\cal R} = 0~.
\end{equation}
Without taking into account the existence of a $\delta$-function in (\ref{cal-r}), one can obtain the harmonic solution,
\begin{equation}
{\cal R} \sim e^{\pm iy} ~,
\end{equation}
which is valid in the whole space of the new variable (\ref{y}), $-\infty \leq y \leq \infty$, including the horizon $y \to -\infty ~(r\to 1)$.

To find correct boundary conditions at the BH horizon, $f (r) = 0$, let us use $f(r)$ as an independent variable in the region $2M \le r \le \infty$ ($0 \le f \le 1$) and rewrite (\ref{R''}) in the form:
\begin{equation} \label{R(f)}
f^2(1-f)^4 R'' + f(1-f)^3(1-3f) R' -\left\{4M^2C + f(1-f)^2\left[l(l+1)+ 1-f\right] \right\} R = 0~,
\end{equation}
where primes denote derivatives with respect to $f$. Keeping the terms with the derivatives and separation constant, which can be large close to the horizon, this equation takes the asymptotic form:
\begin{equation} \label{R(f=0)}
f^2R'' + f R' - 4M^2C R = 0~. ~~~~~ (f \to 0)
\end{equation}
We need to find regular solution to this equation at $f = 0$.

At first let us consider the case with zero separation constant and write the equation (\ref{R(f=0)}) in the form known from the two dimensional potential problem:
\begin{equation} \label{R(C=0)}
R'' + \frac 1f R' = \frac 1f \left(f R'\right)' = 0~.
\end{equation}
We want to emphasize that the famous logarithmic solution to this equation,
\begin{equation} \label{ln}
R \sim \ln \left| \frac 1f \right|~,
\end{equation}
corresponds to a point-like source, $\delta (f)$, and does not represent the solution to (\ref{R(C=0)}) without delta function at the right side, i.e. the regular solution at $f = 0$ is not given by (\ref{ln}) but has the form:
\begin{equation}
R(r)_{C = f = 0} = const ~.
\end{equation}

Another example of misunderstandings with singular coordinate transformations is the appearance of the fictitious delta function at the origin in the definition of the Laplace equation in spherical coordinates \cite{Khel-Nad,Can-Khe}. It is known that regular solutions to the Laplace equation in the Cartesian coordinates,
\begin{equation} \label{Laplace}
\left( \partial_x^2 + \partial_y^2 + \partial_z^2 \right){\cal \varphi } = 0 ~,
\end{equation}
are given by exponential functions of the form
\begin{equation}
{\cal \varphi } \sim e^{\pm [i (ax + by) + \sqrt {a^2 + b^2}z]}~,
\end{equation}
were $a$ and $b$ are integration constants \cite{Jackson}. This solution is regular everywhere and at the origin is constant. After rewriting the Laplace equation (\ref{Laplace}) in spherical coordinates, which is analogous to the singular transformation (\ref{tortoise}), one can obtain fictitious singular solutions in the form of the Newtonian potential,
\begin{equation}
{\cal \varphi } \sim \frac 1r ~.
\end{equation}
However, this solution requires the existence of a delta-function at the origin and does not corresponds to the regular boundary conditions there.

Using these observations we see that the equation (\ref{R(f=0)}) has a regular at $f=0$ solution only for the positive separation constant $C$,
\begin{equation} \label{R->0}
R(r)|_{f \to 0} \sim f^{2M\sqrt C} ~. ~~~~~(C > 0)
\end{equation}
The solution to (\ref{R(f=0)}) with the negative $C$,
\begin{equation} \label{C<0}
R(r)|_{f \to 0} \sim e^{i 2M\sqrt {-C} \ln |f|} ~, ~~~~~(C < 0)
\end{equation}
which is used as a boundary solution in (\ref{solution-f=0}), does not obey (\ref{R''}), since it leads to the extra delta-like source term at the BH horizon.

Note that from the boundary solution (\ref{R->0}), which we want to use below, follows that
\begin{equation}
R(r)_{f \to 0} \rightarrow 0 ~,
\end{equation}
i.e. the probability for particles to cross the horizon tends to zero and when the used approximations are valid the infalling particles are absorbed by a BH or can be reflected from its horizon. In this case the information about quantum particles does not disappear for a distance observer, which can solve the BH information paradox. However, for the cases of massive particles (with the extra term in the effective potential (\ref{Veff})), or the large angular momentum case, when the approximation
\begin{equation}
f~ l(l+1) \ll 4M^2C  ~~~~~ (f \to 0)
\end{equation}
is not applicable, in principle the last term in (\ref{R(f=0)}) can change the sign and probably one can obtain the harmonic solution (\ref{C<0}), corresponding to the matter crossing the BH horizon. This issue needs further investigations.

Now let us look for the solution to (\ref{R(f)}) for the regular boundary condition (\ref{R->0}), using the method of Frobenius:
\begin{equation} \label{R=}
R (f) = \sum_{i}^{\infty} a_{i} f^{i}~, ~~~~~~~~(i = 1, ...,\infty)
\end{equation}
where $a_{i}$ are some constants. The equation (\ref{R(f)}) reduces to the algebraic system for the coefficients of $f^i$, which will be satisfied when the coefficients of each $f^i$ becomes zero. The first equation of this system ($i = 1$),
\begin{equation} \label{a}
f(1-f)^3(1-3f)a_1 -\left\{4M^2C + f(1-f)^2 [l(l+1)+ 1-f] \right\}a_1f \approx \left(1 - 4M^2 C \right)a_1f = 0 ~,
\end{equation}
shows that the separation constant in the system (\ref{T''})-(\ref{R''}) indeed is positive \cite{Qin},
\begin{equation} \label{k}
C \approx \frac {1}{4M^2} > 0 ~.
\end{equation}
In this case from (\ref{T''}) we find the real-valued exponential solution,
\begin{equation} \label{T}
T(t)= T_0 e^{\pm t/2M} ~,
\end{equation}
i.e. in the Schwarzschild coordinates of a distant observer the wave function (\ref{psi=RT}) contains the real exponential factor:
\begin{equation} \label{psi=}
\psi(t,r) \sim \frac {e^{\pm t/2M}}{r} \sum_{i}^{\infty} a_{i} \left(1 - \frac {2M}{r}\right)^{i} ~.
\end{equation}
The constants $a_i$ satisfy the relations \cite{Qin},
\begin{equation}
\frac{a_{i+1}}{a_i}\big|_{i \to \infty} \to 2~,
\end{equation}
what means that the condition of convergence of the radial wave function (\ref{R=}),
\begin{equation}
\frac{a_{i+1} f^{i+1}}{a_i f^i}\big|_{i \to \infty} < 1 ~,
\end{equation}
is valid for the region
\begin{equation} \label{f<1/2}
0 < f < \frac 12~. ~~~~~(2M <r < 4M)
\end{equation}
Thus the solution (\ref{psi=}) can be used in the interesting for us layer between the Schwarzschild and photons spheres,
\begin{equation} \label{2M<r<3M}
2M < r < 3M ~.
\end{equation}

The real-valued function (\ref{psi=}) is very different from the familiar internal \cite{BH-wave} and external \cite{BH-out} periodic-in-time solutions ($\sim e^{i\omega t}$) for scalar particles in the space (\ref{Schwarzschild}). Another difference is that in our case the condition $r \to 2M$ influences only the spatial component of the wave function (\ref{psi=RT}) and never the time component. In semi-classical quantum mechanics a wave function with the complex phase, like (\ref{psi=}), usually is the signature of a tunneling processes through a potential barrier, i.e. the penetration of particles through the BH horizon cannot occur classically.

The similar to (\ref{psi=}) real-valued solutions to the Klein-Gordon equation in Schwarzschild's coordinates was obtained in \cite{Tunneling}, were it was nevertheless assumed that classical geodesics are extendable across the horizon, while the complex phase was connected with the creation of particles by the gravitational field of BHs. But to regularize the propagator of scalar field the singular point at the horizon, which shows that classical particles are stopped from entering the BH, was removed by the introduction of the infinitesimal integration contours around the pole in the integral for the 'tortoise' coordinate (\ref{tortoise}).

Now we want to show that the exponential enhancement (\ref{T}) of amplitudes of waves passed the edge of a BH can be used to explain appearance of energetic GWs, GRBs and FRBs for a distant observer. To study the BH strong lensing it is necessary to investigate isotropic geodesics in the region (\ref{2M<r<3M}). In general, trajectories of classical particles can be considered as the geometrical-optical limiting case (eikonal approximation) of a wave movement \cite{Gold}. It is known that the Klein-Gordon wave functions associated with the classical motion formally obey the relativistic Hamilton-Jacobi equation written for the same system \cite{Motz}. Indeed, the scalar wave function (\ref{Phi}) can be expressed in terms of an amplitude and phase,
\begin{equation} \label{Phi=rho}
\Phi = A e^{iS}~,
\end{equation}
where $S(x^\nu)$ is the Hamilton principal function, which usually is used in the definition of momentum,
\begin{equation}
p^\nu \sim \partial^\nu S ~.
\end{equation}
Then the Klein-Gordon equation (\ref{wave}) gives the system of equations:
\begin{eqnarray} \label{KG-system}
A \Box S + 2 \partial_\nu S \partial^\nu A = 0~, \nonumber \\
\Box A - A \partial_\nu S \partial^\nu S = 0~.
\end{eqnarray}
In the weak gravitational field, in the short wavelength limit, and if one neglects variations of the amplitude:
\begin{equation}
f \to 1~, ~~~~~ \Box S \to 0 ~, ~~~~~ \partial^\nu A \to 0 ~.
\end{equation}
In this approximation from (\ref{KG-system}) follows that the eikonal phase (Hamilton's principal function) can be written as:
\begin{equation} \label{S}
S \sim p_\nu x^\nu~,
\end{equation}
and that $p_\nu$ obeys the  massless relativistic Hamilton-Jacobi (geodesic) equation,
\begin{equation}
g_{\mu\nu}p^\mu p^\nu = 0~.
\end{equation}

Far from a BH a free particle equation has the usual flat wave solution,
\begin{equation} \label{Phi_infty}
\Phi_\infty = A e^{\pm ip_\nu x^\nu} ~,
\end{equation}
but close to its horizon variations of $A$ in (\ref{Phi=rho}) cannot be neglected, since the time dependent factor of the wave function obtains the form (\ref{T}). The time-dependence of amplitudes of the scattered waves in the Schwarzschild field means the apparent non-conservation of energy from the point of view of a distant observer, which is the result of neglecting back-reaction of particles on the Schwarzschild metric in our analysis of the scattering problem. One can still use the flat wave function (\ref{Phi_infty}), but with the time dependent energy in (\ref{S}), $p_0 = p_0(t)$. Indeed, inserting
\begin{equation} \label{A}
A_{r \to 2M} \sim e^{\pm t/2M}
\end{equation}
into the system (\ref{KG-system}) we find
\begin{eqnarray} \label{KG-A}
\partial_t p_0 \pm \frac 1M p_0 = 0~, \nonumber \\
g_{\mu\nu}p^\mu p^\nu - \frac {1}{4M^2}= 0~.
\end{eqnarray}
This means that the energy of incident flat waves is changed with time as,
\begin{equation} \label{E(t)}
p_0(t) \sim e^{\pm t/M}~,
\end{equation}
and also that close to the horizon photons acquire the effective mass, $\sim 1/2M$. The time interval that photons need to make a loop of the radius $r \sim 2M$ around the BH is of the order of
\begin{equation}
\int_0^{2\pi} r d\theta \sim 4\pi M ~.
\end{equation}
Thus the energy of outgoing photons, due to the exponential amplification factor from (\ref{A}), can reach the huge value,
\begin{equation}
p_0^{ref} \sim e^{4\pi} p_0 \approx 3\cdot 10^5 p_0 ~,
\end{equation}
and a distant observer can receive burst-like outgoing waves of the form (\ref{Phi_infty}), where now $p_0$ is replaced by $p_0^{ref}$.

To write appropriate geodesic equation which describes strong lensings we want to remind also the models of classical interpretation of the Schr\"{o}dinger wave function, where variations of the wave function amplitude are taking into account by introduction of so called quantum potential \cite{Schro,Gro}. For example, in the thermodynamic approach \cite{Gro}, the variation of the distribution density (wave amplitude) of a moving particle is modeled by a resistance of the ensemble in the form of the heat flow. For the BH case, appearance of the heat flow close to the Schwarzschild horizon is analogous to the 'firewall' conjecture, the existence of energetic curtain at the event horizon \cite{Firewall}, as if BHs in the range (\ref{2M<r<3M}) have effective quantum 'atmospheres' \cite{BH-atmosphere}. It is worth to mention also the models were an observer at infinity may detect the emitted particles which extract energy from the rotating BH due to the spontaneous polarization process inside its ergosphere \cite{ergosphere}.

Using these analogies, in the quasi-classical approximation, one can still describe trajectories of particles close to a BH horizon by ordinary geodesic equations, but with extra potential, which takes into account variations of the amplitude (\ref{T}). Relativistic invariant vacuum energy in the General Relativity is known under the name of cosmological constant, $\Lambda$. Solution of the spherically symmetric Einstein equations with the cosmological term is well known, provided that the factor $f(r)$ is replaced in (\ref{Schwarzschild}) by the function:
\begin{equation} \label{F(r)}
F(r) = 1 - \frac {2M}{r} - \frac {\Lambda}{3} r^2~.
\end{equation}
In our case the constant $\Lambda$ should be fixed in the way which enables us to use Klein-Gordon equation (\ref{psi}) close to the BH horizon and at the same time cancels apparent energy non-conservation for a distant observer.

Inserting the function (\ref{F(r)}) into the system (\ref{T''})-(\ref{R''}), instead of $f(r)$, and expanding the solution of the modified equation (\ref{R(f)}) as the series of $F(r)$, the equation (\ref{a}) takes the form:
\begin{eqnarray} \label{a_1}
F(1-F)^3(1-3F)a_1 -\{4M^2C + F(1-F)^2 [l(l+1)+ 1-F] \}a_1F = \\
= [(1-F)^3(1-4F)- 4M^2C - F(1-F)^2l(l+1)]a_1F = 0 ~.\nonumber
\end{eqnarray}
For $F \to 0$ this equation, as for ordinary Schwarzschild case, gives (\ref{k}). The second real valued solutions of (\ref{a_1}) for small multipoles, $l\le 3$, is
\begin{equation} \label{F}
F \approx 2 ~.
\end{equation}
We require validity of (\ref{a_1}) at the horizons of $F(r)$ and $f(r)$ simultaneously, i.e. at $F \to 0$ and at $F \to -4M^2\Lambda /3$. Then (\ref{F}) leads to
\begin{equation} \label{Lambda}
\frac \Lambda 3\approx - \frac {1}{2 M^2}~.
\end{equation}
So the effective cosmological constant $\Lambda$ in our model is negative and
\begin{equation} \label{F=}
F(r) = 1 - \frac {2M}{r} + \frac {r^2}{2 M^2}~,
\end{equation}
what corresponds to the effective AdS space for classical particles moving inside the shell (\ref{2M<r<3M}) where we want to study geodesics.

It is known that the classical geodesic equations can be obtained using the Lagrangian \cite{Chandra,BH-AdS},
\begin{equation} \label{L}
L \left(x^\nu\right) = \frac 12 g_{\alpha\beta}u^\alpha u^\beta ~,
\end{equation}
where
\begin{equation}
u^\alpha = \frac {dx^\alpha}{ds}
\end{equation}
denotes the 4-velocity and $ds$ is the proper time. From (\ref{F=}) and (\ref{L}) we find the expressions for two conserved components of 4-momentum in the space-time (\ref{Schwarzschild}):
\begin{eqnarray}\label{p_t,p_phi}
\frac{\partial L(x^\nu)}{\partial u^t} &=& u_t F(r) = E~, ~~~~~~~ \left(\theta = \frac \pi 2\right) \nonumber \\
\frac{\partial L(x^\nu)}{\partial u^\phi} &=& u_\phi r^2 = J~,
\end{eqnarray}
where $E$ and $J$ denote the waves energy and angular momentum and
\begin{equation}
u_t = \frac {dt}{ds}~, ~~~~~ u_\phi = \frac {d\phi}{ds}
\end{equation}
are the 4-velocity components. Using (\ref{p_t,p_phi}) the Lagrangian (\ref{L}) can be write in the following form:
\begin{equation} \label{radial}
2L = F u_t^2 - \frac {u_r^2}{F} - r^2 u_\phi^2 = \epsilon~, ~~~~~ (u_\theta = 0)
\end{equation}
where we can consider that $\epsilon = 1$ for massive particles and $\epsilon = 0$ for isotropic geodesics \cite{Chandra,BH-AdS}. The expression (\ref{radial}) represents the first integral of the last independent geodesic equation,
\begin{equation} \label{p_r}
u_r^2 = E^2 - F(r)\left(\epsilon + \frac {J^2}{r^2}\right) = E^2 - U^2~,
\end{equation}
with
\begin{equation}
u_r = \frac {dr}{ds}~.
\end{equation}
The effective potential in (\ref{p_r}),
\begin{equation} \label{U}
U^2 = \left(1 - \frac {2M}{r} + \frac {r^2}{2 M^2}\right)\left(\epsilon + \frac {J^2}{r^2}\right) = U_{Sch}^2 + \epsilon \frac {r^2}{2 M^2} + \frac {J^2}{2 M^2}~,
\end{equation}
differs from the ordinary Schwarzschild's one,
\begin{equation} \label{U_Sch}
U_{Sch}^2 = \left(1 - \frac {2M}{r}\right)\left(\epsilon + \frac {J^2}{r^2}\right)~,
\end{equation}
by the nonnegative terms, $\epsilon r^2/2 M^2$ and $J^2/2M^2$, thus its zero is shifted inside the Schwarzschild sphere \cite{BH-AdS},
\begin{equation} \label{F-horizon}
r = \frac{2}{\sqrt {-\Lambda}} \sinh \left[ \frac 13 {\rm arsinh} \left(3M\sqrt {-\Lambda} \right) \right] \approx 1.1 M~.
\end{equation}

\begin{figure}
\begin{center}
\resizebox{0.45\textwidth}{!}{\includegraphics{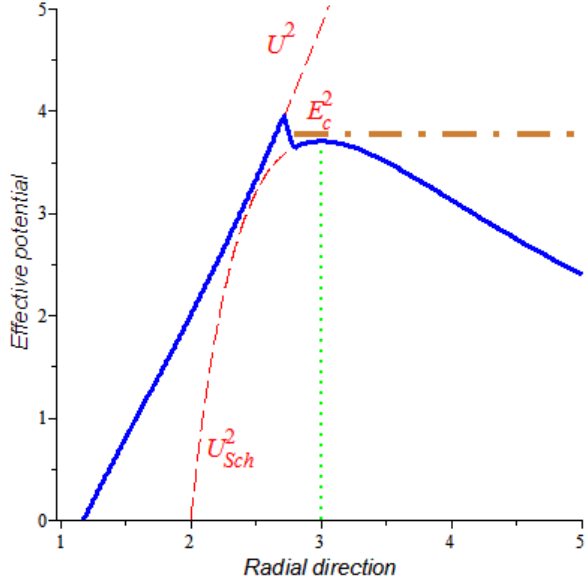}}
\caption{The effective potential for non-radial waves.}
\label{fig:1}
\end{center}
\end{figure}

FIG 1. displays the effective potential in our model (bold blue line) for $J=10$ and $M=1$. Outside the photon sphere (where $\Lambda = 0$ and $\epsilon = 0$) the effective potential is given by $U^2_{Sch}$. It is clear that a non-radial isotropic geodesics (dashed line) with the energy above or equal to the critical value,
\begin{equation}
E^2_c = \frac {J^2}{27 M^2}~,
\end{equation}
could cross the photons sphere \cite{Chandra, BH-AdS}. According to our model, inside the shell between the Schwarzschild and photon spheres (\ref{2M<r<3M}) waves occur in the effective AdS space, where photons acquire effective mass and $\epsilon = 1$. In this region the effective potential (\ref{U}) forms the second higher barrier and the actual impact factor changes the sign. So waves could stay in the minimum between two pics of $U^2(r)$ looping on some bound orbit, or reflect back (FIG. 1). The situation reminds one of the Meissner effect when photons obtain effective masses and are ejected from the superconducting region.

Inside the photon sphere the energy of photons will change according to (\ref{E(t)}), where $t$ is the time interval photons spending there. Then reflected photons with the increased energy could form burst-like short signals from the BH edge, mainly in gamma and radio frequencies which could escape surrounding the BH dust clouds. In the case of GWs dust is not a problem and there is chance to observe also relativistic lensing. So additionally we need to explore the obtained from (\ref{p_t,p_phi}) and (\ref{p_r}) radial equation:
\begin{equation} \label{dr/dt}
\left(\frac{dr}{dt}\right)^2 = F^2(r) \left[27 M^2\frac{F(r)}{r^2} - 1\right] ~.
\end{equation}
Inside the shell (\ref{2M<r<3M}) we can use the approximation
\begin{equation}
F(r) \approx \frac {r^2}{2M^2}~.
\end{equation}
For the inspiraling towards the center GWs ($dr/dt < 0$) the equation (\ref{dr/dt}) gives:
\begin{equation}
r \approx \frac {15 M^2}{27t + 5M} \approx 3M - 16 t~,
\end{equation}
where the integration constant is fixed from the condition
\begin{equation}
r(t=0) = 3 M~.
\end{equation}
From (\ref{F=}) and (\ref{p_t,p_phi}) for the instant frequency we find:
\begin{equation} \label{omega}
\omega = \frac{d\phi}{d t} = \frac{3\sqrt 3 M}{r^2} F(r)  = \frac {14}{5 M} + \frac {9}{5M^2}t~.
\end{equation}
When the effective impact parameter changes sign, looping GWs will start deflecting and a distant observer will receive the periodic chirp like signals with the exponential amplitudes (\ref{T}) and increasing frequency (\ref{omega}). Then the expression for the timing of the resulted strain,
\begin{equation} \label{h(t)}
h \sim A e^{t/2M}\sin \omega t + B e^{-t/2M}\cos \omega t~,
\end{equation}
where $A$ and $B$ are some constants, can be described by a linear superposition of several quasi-normal modes of perturbed BH horizon by GWs \cite{Ringdown} and will imitate the LIGO signals \cite{GW}.

To conclude, in this paper we have reanalyzed boundary conditions for the equations of matter particles at a BH horizon and found the real-valued exponentially time-dependent solutions. While quantum wave functions and isotropic geodesics are continuous below the photons sphere, $r < 3M$, for a distant observer receiving signals only from the region, $r > 2M$, these exponential enhancement (decay) of amplitudes will be visible as if the gravitationally lensed waves receiving (giving) the energy from (to) the BH horizon. To model isotropic geodesics in the strong gravitational field between the Schwarzschild and photon spheres we have introduced the effective negative cosmological constant with the value that guaranties validity of the eikonal approximation in this region. Then we found that part of the incident waves crossing the effective AdS space below the photon sphere can be amplified and reflected and this mechanism can explain some GWs, GRBs and FRBs.



\begin{thebibliography}{99}
\bibitem{GW} LIGO Scientific and Virgo Collaborations,
            Phys. Rev. Lett. {\bf 116} (2016) 061102, arXiv: 1602.03837 [gr-qc];
            Phys. Rev. Lett. {\bf 116} (2016) 241103, arXiv: 1606.04855 [gr-qc].

\bibitem{GRB} T. Piran,
             Phys. Rept. {\bf 314} (1999) 575, arXiv: astro-ph/9810256; \\
              P. Meszaros,
             Rept. Prog. Phys. {\bf 69} (2006) 2259, arXiv: astro-ph/0605208; \\
              P. Kumar and B. Zhang,
             Phys. Rept. {\bf 561} (2014) 1, arXiv: 1410.0679 [astro-ph.HE].

\bibitem{FRB} D.R. Lorimer, M. Bailes, M.A. McLaughlin, D.J. Narkevic and F. Crawford,
             Science {\bf 318} (2007) 777, arXiv: 0709.4301 [astro-ph]; \\
              D. Thornton, {\it et al.},
             Science {\bf 341} (2013) 53, arXiv: 1307.1628 [astro-ph.HE]; \\
              E.F. Keane, {\it et al.},
             Nature {\bf 530} (2016) 453, arXiv: 1602.07477 [astro-ph.HE].

\bibitem{Lens} P. Schneider, J. Ehlers and E.E. Falco,
             {\it Gravitational Lenses} (Springer, Berlin 1992); \\
              P. Schneider, C. Kochanek and J. Wambsganss,
             {\it Gravitational Lensing: Strong, Weak and Micro} (Springer, Berlin 2006).

\bibitem{Darwin} C. Darwin,
                Proc. R. Soc. London {\bf A 249} (1959) 180; ibid. {\bf 263} (1961) 39; \\
                 R. Atkinson,
                The Astronomical Journal {\bf 70} (1965) 517.

\bibitem{StrLens} S. Frittelli, T.P. Kling and E.T. Newman,
                Phys. Rev. {\bf D 61} (2000) 064021, arXiv: gr-qc/0001037; \\
                  V. Perlick,
                Living Rev. Rel.  {\bf 7} (2004) 9.

\bibitem{Virb} K.S. Virbhadra and G.F.R. Ellis,
              Phys. Rev. {\bf D 62} (2000) 084003, arXiv: astro-ph/9904193; \\
               K.S. Virbhadra,
              Phys. Rev. {\bf D 79} (2009) 083004, arXiv: 0810.2109 [gr-qc].

\bibitem{Gog} M. Gogberashvili,
             arXiv: 1712.02637 [gr-qc].

\bibitem{Mar} D. Marolf,
             Rep. Prog. Phys. {\bf 80} (2017) 092001, arXiv: 1703.02143 [gr-qc].

\bibitem{Chandra} S. Chandrasekhar,
                {\it The Mathematical Theory of Black Holes}
                (Clarendon, New York 1983).

\bibitem{Star} A.A. Starobinskii,
              Sov. Phys. JETP {\bf 37} (1973) 28.

\bibitem{Matz} R.A. Matzner,
              J. Mat. Phys. {\bf 9} (1968) 163.

\bibitem{Khel-Nad} A. Khelashvili and T. Nadareishvili
                 Am. J. Phys. {\bf 79} (2011) 668, arXiv: 1009.2694 [quant-ph];
                 Phys. Part. Nucl. Lett. {\bf 12} (2015) 11, arXiv: 1502.04008 [hep-th].

\bibitem{Can-Khe} Y.C. Cantelaube and A.L. Khelif,
                 J. Math. Phys. {\bf 51} (2010) 053518.

\bibitem{Jackson} J.D. Jackson,
                 {\it Classical  Electrodynamics} (Wiley, NY 1999).

\bibitem{Qin} Y.-P. Qin,
             Sci. China: Phys. Mech. Astron. {\bf 55} (2012) 381.

\bibitem{BH-wave} T. Damour and R. Ruffini,
                 Phys. Rev. {\bf D 14} (1976) 332; \\
                  S. Sannan,
                 Gen. Rel. Grav. {\bf 20} (1988) 239.

\bibitem{BH-out} E. Elizalde,
                Phys. Rev. {\bf D 36} (1987) 1269.

\bibitem{Tunneling} K. Srinivasan and T. Padmanabhan,
                  Phys.Rev. {\bf D 60} (1999) 024007, arXiv: gr-qc/9812028.

\bibitem{Gold} H. Goldstein,
              {\it Classical Mechanics} (Addison-Wesley, New York 1950).

\bibitem{Motz} L. Motz and A. Selzer,
              Phys. Rev. {\bf 133} (1964) B1622.

\bibitem{Schro} D. Bohm,
               Phys. Rev. {\bf 85} (1952) 166; \\
                E. Nelson,
               {\it Quantum Fluctuations} (Princeton Univ. Press, Princeton 1985); \\
                M.J.W. Hall and M. Reginatto,
               J. Phys. {\bf A 35} (2002) 3289, arXiv: quant-ph/0102069.

\bibitem{Gro} G. Gr\"ossing,
             Phys. Lett. {\bf A 372} (2008) 4556, arXiv: 0711.4954 [quant-ph];
             Found. Phys. Lett. {\bf 17} (2004) 343, arXiv: quant-ph/0311109; \\
              M. Gogberashvili,
             Int. J. Theor. Phys. {\bf 50} (2011) 2391, arXiv: 1008.2544 [gr-qc].

\bibitem{Firewall} A. Almheiri, D. Marolf, J. Polchinski and J. Sully,
                  JHEP {\bf 1302} (2013) 062, arXiv: 1207.3123 [hep-th]; \\
                   S.L. Braunstein,
                  Phys. Rev. Lett. {\bf 110} (2013) 101301, arXiv: 0907.1190 [quant-ph].

\bibitem{BH-atmosphere} S.B. Giddings,
                       Phys. Lett. {\bf B 754} (2016) 39, arXiv: 1511.08221 [hep-th].

\bibitem{ergosphere} R. Penrose,
                    Riv. Nuovo Cimento {\bf 1} (1969) 252; \\
                     T. Piran and J. Shaham,
                    Phys. Rev. {\bf D 16} (1977) 1615.

\bibitem{BH-AdS} N. Cruz, M. Olivares and J.R. Villanueva,
                Class. Quant. Grav. {\bf 22} (2005) 1167, arXiv: gr-qc/0408016.

\bibitem{Ringdown} K.D. Kokkotas and B. G. Schmidt,
                  Living Rev. Rel. {\bf 2} (1999) 2, arXiv: gr-qc/9909058; \\
                   R.A. Konoplya and A. Zhidenko,
                  Rev. Mod. Phys. {\bf 83} (2011) 793, arXiv: 1102.4014 [gr-qc].

\end{thebibliography}
\end{document}